\documentclass[pre, twocolumn, showpacs, superscriptaddress]{revtex4}
\usepackage{amsmath}
\usepackage{dcolumn}
\usepackage[xdvi]{graphicx}%
\makeatletter

\def\btt#1{\texttt{\@backslashchar#1}}%
\DeclareRobustCommand\bblash{\btt{\@backslashchar}}%
\makeatother

\newcommand{\bv}[1]{{\boldsymbol #1}}

\begin{document}

\title{Critical scaling near jamming transition for frictional granular particles}

\author{ Michio Otsuki$^1$ and Hisao Hayakawa$^2$}
\affiliation{
$^1$ Department of Physics and Mathematics, Aoyama Gakuin University,
5-10-1 Fuchinobe, Sagamihara, Kanagawa 229-8558, Japan \\
  $^2$ Yukawa Institute for Theoretical Physics, Kyoto University,  Kitashirakawaoiwake-cho, Sakyo-ku, Kyoto 606-8502, Japan}

\begin{abstract}
The critical rheology of sheared frictional granular materials near jamming transition 
is numerically investigated.
It is confirmed that there exist a true critical density
which characterizes the onset of the yield
stress, and two fictitious critical densities
which characterize the scaling laws of rheological properties.
We find the existence of a hysteresis loop between two of the critical
densities for each friction coefficient.
It is noteworthy that the critical scaling law for frictionless jamming transition seems to be still valid even for
frictional jamming despite using fictitious critical density values.

\end{abstract}


\maketitle

\section{Introduction}

Disordered materials such as
granular materials \cite{Jaeger},
colloidal suspensions \cite{Pusey}, emulsions, and foams \cite{Durian}
have rigidity above a critical value of density 
and do not have any rigidity below that value.
This sudden change in rigidity is known as a jamming transition,
which could be
 a key concept for explaining the behavior exhibited by disordered 
 materials at zero temperature \cite{Liu}.
The coordination number changes discontinuously 
at the jamming transition point (point J) for static frictionless spheres
\cite{OHern02,OHern03}.
Critical
scaling laws, similar to those in the continuous phase transition, can
be used to determine rheological properties of granular particles
\cite{Hatano07,Olsson,Hatano08,Tighe,Hatano10,Otsuki08,Otsuki09,Otsuki10}. 
The critical exponents of these scaling laws remain a topic of
current investigations \cite{Hatano07,Olsson,Hatano08,Tighe,Hatano10}. 
For frictionless grains, where
inertial effects in the equations of motion are included, the present
authors \cite{Otsuki08,Otsuki09,Otsuki10} used a mean-field theory to derive a viscosity $\eta$
below point J that satisfies the expression $(\phi_J - \phi)^{-4}$, 
as the packing fraction $\phi$ approaches
the critical jamming fraction $\phi_J$ from below. We believe this
result to be asymptotically valid in the hard core and elastic limits
\cite{Otsuki10}.



On the other hand, recent studies on the jamming transition of frictional grains revealed that elastic moduli, 
coordination number, and density of state
 are strongly affected    
by the introduction of friction
\cite{Silbert02, Zhang, Shudyak,Somfai,Hecke,Henkes,Silbert10}.
However, we still do not know  the details of the quantitative change induced in the rheological properties
of the jamming transition by the introduction of friction between grains.
Moreover, in spite of 
 many studies on the scaling law of the jamming transition 
for sheared frictionless particles, 
there are few such studies on  
the jamming transition of frictional particles.
Thus, we need to clarify whether scaling laws in the vicinity of the jamming transition of frictional grains can be used.

In this paper, we have numerically 
investigated the properties of sheared frictional granular particles
in the vicinity of the jamming transition.
In the next section, the details of our numerical results will be presented.
In Sec. \ref{Setup:sec}, 
we will explain our setup and models.
We will demonstrate the existence of hysteresis loops for 
pressure and shear stress in Sec. \ref{Hysteresis:sec}.
The three critical area fractions for sheared granular materials will be estimated 
in Sec. \ref{Critical:sec}, where one of them is the true critical fraction for the jamming transition and the others are fictitious critical fractions.
In Sec. \ref{Scaling:sec}, we will demonstrate that the scaling relations for
frictionless particles \cite{Otsuki08,Otsuki09}
can be used even for frictional systems by using the fictitious critical fractions.
Finally, we will  discuss and conclude our results 
in Sec. \ref{Discussion:sec}.

\section{Numerical result}
\label{Numerical:sec}

\subsection{Setup of our simulation}
\label{Setup:sec}
Let us consider a two-dimensional frictional granular assembly.
The system includes $N$ grains, each having
an identical mass $m$. The position, velocity, and angular velocity
of a grain $i$ are respectively denoted by
$\bv{r}_i$, $\bv{v}_i$, and $\omega_i$.
Our system consists of grains having 
the diameters $0.7 \sigma_0$, $0.8 \sigma_0$, $0.9 \sigma_0$, 
and $\sigma_0$, where there are $N/4$ for each species of grains.

The contact force $\bv{f}_{ij}$ consists of
the normal part $\bv{f}^{(n)}_{ij}$ and the tangential part $\bv{f}^{(t)}_{ij}$
as $\bv{f}_{ij} = \bv{f}^{(n)}_{ij} + \bv{f}^{(t)}_{ij}$.
The normal contact force $\bv{f}^{(n)}_{ij}$ between the grain $i$
and the grain $j$ is given by
$\bv{f}^{(n)}_{ij} = h^{(n)}_{ij} \Theta(h^{(n)}_{ij})
\Theta (\sigma_{ij} - r_{ij}) \bv{n}_{ij}$,
where $h^{(n)}_{ij}$ and $\bv{n}_{ij}$ are respectively given by
$h^{(n)}_{ij} = k^{(n)} (\sigma_{ij} - r_{ij}) - \eta^{(n)} v^{(n)}_{ij}$
and $\bv{n}_{ij} = \bv{r}_{ij}/|\bv{r}_{ij}|$ for the normal 
elastic constant $k^{(n)}$,
the normal viscous constant $\eta^{(n)}$, 
the diameter $\sigma_i$ of grain $i$,
$\bv{r}_{ij} \equiv \bv{r}_{i} - \bv{r}_{j} $, 
$\sigma_{ij} \equiv (\sigma_i + \sigma_j)/2$ and 
$v^{(n)}_{ij} \equiv (\bv{v}_{i}- \bv{v}_{j}) \cdot \bv{n}_{ij}$
\cite{force:note}. Here, 
$\Theta(x)$ is the Heaviside step function defined by $\Theta(x)=1$
for $x \ge 0$ and $\Theta(x)=0$ otherwise.
Similarly, the tangential contact force
$\bv{f}^{(t)}_{ij}$ between grain $i$
and grain $j$ is given by the equation 
$\bv{f}^{(t)}_{ij} = \min(|h^{(t)}_{ij}|, \mu |\bv{f}^{(n)}_{ij}|) \mathrm{sign} (h^{(t)}_{ij}) \bv{t}_{ij}$,
where $\min(a,b)$ selects the smaller one between $a$ and $b$,
and $h^{(t)}_{ij}$ is given by
$h^{(t)}_{ij} = k^{(t)} u^{(t)}_{ij} - \eta^{(t)} v^{(t)}_{ij}$
with the tangential unit vector 
$\bv{t}_{ij} = (-y_{ij}/|\bv{r}_{ij}|, x_{ij}/|\bv{r}_{ij}|)$.
Here, $k^{(t)}$ and $\eta^{(t)}$ are the elastic and viscous
constants along the tangential direction.
The tangential velocity $v^{(t)}_{ij}$ and the tangential displacement
$u^{(t)}_{ij}$ are respectively given by
$v^{(t)}_{ij} = (\bv{v}_{i}- \bv{v}_{j}) \cdot \bv{t}_{ij}
+ (\sigma_i \omega_i + \sigma_j \omega_j)/2$
and $u^{(t)}_{ij} = \int_{\mathrm{stick}} dt \  v^{(t)}_{ij}$,
where  ``stick'' on the integral indicates that
the integral is performed 
when the condition $|h^{(t)}_{ij}| < \mu |\bv{f}^{(n)}_{ij}|$ or 
 another condition $u^{(t)}_{ij} v^{(t)}_{ij} < 0$
is satisfied \cite{DEM, Hatano09}.

We study the shear stress $S$ and the pressure $P$,
which are respectively given by 
\begin{eqnarray}
S & = &  -\frac{1}{V}\left <    \sum_i^N \sum_{j>i} 
r_{ij,x}
 \left [ f^{(n)}_{ij,y} + f^{(t)}_{ij,y} 
 \right ]
\right > 
\label{S:calc},  \\
P & = & 
\frac{1}{2V} \left < \sum_i^N \sum_{j>i} \bv{r}_{ij} \cdot
 \left [ \bv{f}^{(n)}_{ij} + \bv{f}^{(t)}_{ij} \right ]
\right > ,
\label{P:ex}
\end{eqnarray}
where $V$ is the volume of the system, and
$\left < \cdot \right >$ represents the ensemble average.
Here, we ignore the kinetic parts of $S$ and $P$,
which are respectively given by
$S_{\rm K} =  - \left <    \sum_i^N
p_{i,x} p_{i,y} \right > / (mV)$ and $P_{\rm K} = 
\left <    \sum_i^N
\bv{p}_{i} \cdot \bv{p}_{i}  \right >/ (2mV)$,
because they are significantly smaller than the potential parts in Eqs. (\ref{S:calc}) and (\ref{P:ex})
near the jamming transition point.

In this paper, the shear is imposed 
along the  $y$ direction and macroscopic displacement 
only along the $x$ direction by the following three methods.
The first method is the SLLOD algorithm
under the Lees-Edwards boundary condition \cite{Evans}  which we call ``SL'' for later discussion,
where the time evolution is given by
\begin{eqnarray}
\frac{d \bv{r}_i}{dt} & = & \frac{\bv{p}_i}{m} + \dot \gamma y_i \bv{e}_x,
\label{SLLOD:1} \\
\frac{d \bv{p}_i}{dt} & = & \sum_{j \neq i} \bv{f}_{ij} - 
\dot \gamma p_{i,y} \bv{e}_x,
\label{SLLOD:2}
\end{eqnarray}
with the peculiar momentum $\bv{p}_i = m (\bv{v}_i - \dot \gamma y \bv{e}_x)$
and the unit vector parallel to the $x$-direction $\bv{e}_x$.
In this method, the shear rate $\dot \gamma$ is  a control parameter.

The second method is {\it quasi-static shearing method}, which we call ``QS'' 
\cite{Heussinger09,Vagberg10}.
In this method,
the shear strain $\Delta \gamma$ is applied by an affine transformation
of the position of the particles.
Then, the particles are relaxed under the time evolution equations
\begin{eqnarray}
\frac{d \bv{r}_i}{dt} & = & \frac{\bv{v}_i}{m}, \\
\frac{d \bv{v}_i}{dt} & = & \sum_{j \neq i} \bv{f}_{ij}
\end{eqnarray}
until the kinetic energy per particles becomes lower than
a threshold value $E_{\rm th}$.
Then, we repeat applying the shear and the relaxation process.
Here, we chose $\Delta \gamma = 10^{-6}$ and $E_{\rm th} = 10^{-7} k^{(n)}\sigma_0^2$,
which are small enough not to influence our results.
This method is expected to correspond to the low shear limit of the SL method.

The third method is {\it stress control method}, which we call ``SC''.
Here, the time evolution equations are given by Eqs. \eqref{SLLOD:1} 
and \eqref{SLLOD:2},
under the time evolution of the shear rate $\dot \gamma$
\begin{eqnarray}
\frac{d \dot \gamma }{dt} & = & \frac{S_0 - S}{Q}
\end{eqnarray}
with the relaxation constant $Q = 10^4k^{(n)}/\sigma_0$
\cite{Evans}.
Here, we choose $S_0 = 10^{-7}$, which is small enough that
the results do not depend on the value of $S_0$.

In this paper, we mainly use
the SL method for the analysis. 
We also examine 
the QS and SC methods
in Figs. \ref{f_fig}, \ref{P_QS}, and \ref{f_size} in Sec. II. C to determine the true jamming density.

\subsection{Hysteresis in rheological properties}
\label{Hysteresis:sec}

First, let us consider  rheological properties of sheared granular particles.
Here, we study the shear stress $S$ and the pressure $P$ by using the SL method.
We find that the shear stress $S$ or the pressure $P$
suddenly changes for a finite friction constant $\mu$ at a critical value of $\dot \gamma$ near the jamming point.
This might be related to the existence of a hysteresis loop
for frictional granular materials
\cite{Zhang,Hysteresis}.
To verify the existence of the hysteresis loop,
we first vary the shear rate $\dot \gamma$ 
from $\dot \gamma_0 = 5.0 \times 10^{-4} \sqrt{k^{(n)}/m}$
to sequentially decreasing values as 
$\dot \gamma = \dot \gamma_0,  a^{-1} \dot \gamma_0,
a^{-2} \dot \gamma_0, \cdots,  a^{-N_s} \dot \gamma_0$
with the rate of change $a$ and the number of the step $N_s = 2 \ln 10 / \ln a$,
where the smaller $\dot \gamma$ is $ 5.0 \times 10^{-6} \sqrt{k^{(n)}/m}$.
We call this process the {\it decreasing process}.
Next, we vary the shear rate from an initial values of 
$a^{-N_s} \dot \gamma_0$
to sequentially increasing values given by
$\dot \gamma = a^{-N_s} \dot \gamma_0,
a^{-N_s - 1} \dot \gamma_0,  \cdots,  a^{-2} \dot \gamma_0,
a^{-1} \dot \gamma_0, \dot \gamma_0$.
We call this process the {\it increasing process}.
Here, we vary the shear rate so  the data as to be apart in the logarithmic plot.
For each value of shear rate $\dot \gamma$,
the system attains a quasi-steady state 
after the waiting time is larger than  $\tau_{\rm w} = 0.4 \dot \gamma^{-1}$,
and we have measured the shear stress $S$
and the pressure $P$ during the time $\tau_{\dot \gamma} = 2.0 \dot \gamma^{-1}$
after the waiting time is equal to $\tau_{\rm w}$.

\begin{figure*}
\begin{center}
\includegraphics[width=0.7 \linewidth]{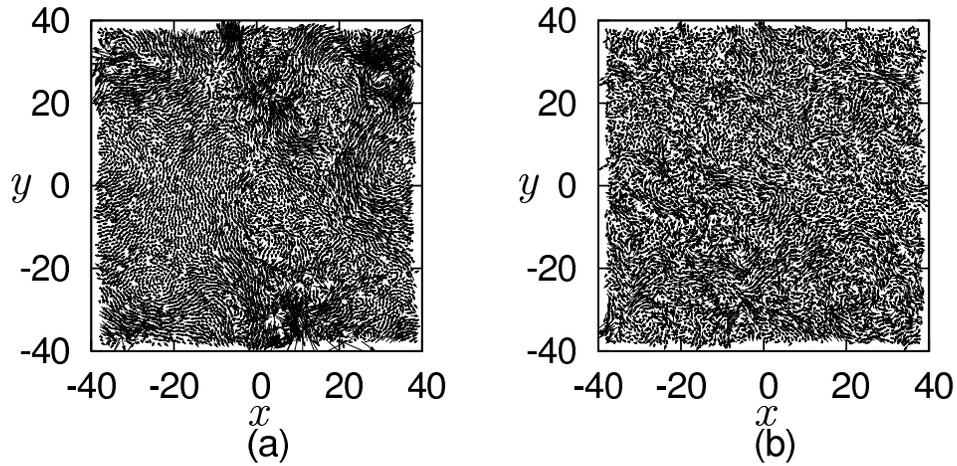}
\caption{ 
 Snapshot of the scaled peculiar momentum field $\bv{p}_i / \sqrt{mT}$
 for the {\it decreasing process} (a) and the {\it increasing process} (b)
  for $\mu = 2.0$, $\phi=0.795$, $N=8000$, $a = 10^{0.1}$,
    and $\dot \gamma = 7.9 \times 10^{-6}\sqrt{k^{(n)}/m}$.
  }
\label{Branch_snap}
\end{center}
\end{figure*}

As expected, there exists a hysteresis loop in frictional jammed systems 
under this setup.
Figure \ref{Branch_snap} shows the
scaled peculiar momentum $\bv{p}_i / \sqrt{mT}$
with
granular temperature $T = \sum_i | \bv{p}_i|^2 / (2mN)$
in both the {\it decreasing process} (Fig.1 (a)) and the {\it increasing process} (Fig.1 (b))
for $\mu = 2.0$, $\phi=0.795$, $N=8000$, $a=10^{-0.05}$ 
  and $\dot \gamma = 7.9 \times 10^{-6}\sqrt{k^{(n)}/m}$.
In the {\it decreasing process} 
(Fig. \ref{Branch_snap}(a)),
the motion of grains is localized and the characteristic length scale 
is significantly 
larger than the grain size. In the {\it increasing process},
as shown in Fig. \ref{Branch_snap}(b), the heterogeneity of grain motion is rather suppressed. 
It should be noted that the coordination number $Z$ in 
Fig. \ref{Branch_snap}(a) is larger than $3$,
whereas $Z$ in Fig. \ref{Branch_snap}(b) is less than $2$.
This difference in the coordination numbers might be related to 
the hysteretic behavior of frictional granular particles.

\begin{figure*}
\begin{center}
\includegraphics[width=0.7 \linewidth]{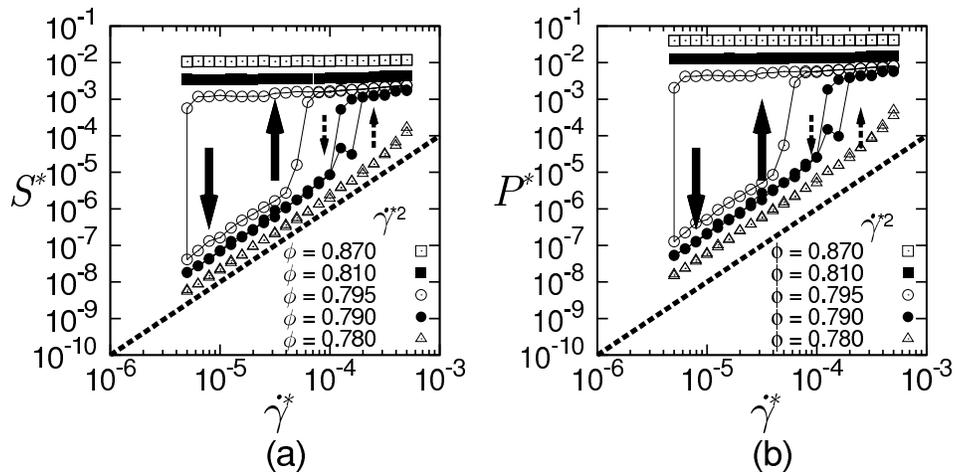}
\caption{ 
  (a) Scaled shear stress $S^* = S /(k^{(n)} \sigma_0^{-1})$ as a function of 
  scaled shear rate $\dot \gamma ^* = \dot \gamma \sqrt{k^{(n)}/m}$ for various 
  values of packing
  fraction $\phi$ for $\mu = 2.0$, $N=8000$, and $a = 10^{0.1}$. \\
  (b) Scaled pressure $P^* = P /(k^{(n)} \sigma_0^{-1})$ as a function of $\dot \gamma^*$ for various values of packing
  fraction $\phi$ for $\mu = 2.0$, $N=8000$, and $a = 10^{0.1}$. 
}
\label{rheology}
\end{center}
\end{figure*}

The shear stress $S$ and the pressure $P$  exhibit clear hysteresis loops,
as shown in Fig. \ref{rheology}.
In Fig.\ref{rheology}(a), we plot the shear stress $S$ 
in a quasi-steady state of each shear rate $\dot \gamma$ 
with $\mu=2.0$, $N=8000$, and $a = 10^{0.1}$ 
for packing fraction values of $\phi=0.870, 0.810, 0.795, 0.790$,
and $0.780$.
For highly packed systems such as $\phi=0.870$ or $0.810$, 
$S$ or $P$ becomes a constant
in a weak shear limit ($\dot \gamma \to 0$),
which implies the existence of the yield stress.
In contrast, $S$ and $P$ are proportional to $\dot \gamma^2$
for a relatively low packed system at $\phi = 0.780$.
It should be noted that  
$S$ and $P$ are independent of the process for both limits, 
as in the case of frictionless particles 
\cite{Hatano07,Hatano08,Otsuki08,Otsuki09}.
However,  $S$ and $P$ depend on the process
 for intermediate packed systems at $\phi=0.795$ and $0.790$,
 and thus, hysteresis loops appear in this region.
For $\phi = 0.795$, $S$ and $P$ 
suddenly decrease at around $\dot \gamma = 5.0 \times 10^{-6}\sqrt{k^{(n)}/m}$
in the {\it decreasing process},
but increases  around at $\dot \gamma = 5.0 \times 10^{-5}\sqrt{k^{(n)}/m}$ 
in the {\it increasing process}
as shown in Fig. \ref{rheology}(a).
The upper branch in the {\it decreasing process}
is called the {\it solid branch}
and the lower branch in the {\it increasing process},
the {\it liquid branch}.


\begin{figure}
\begin{center}
\includegraphics[width=8cm]{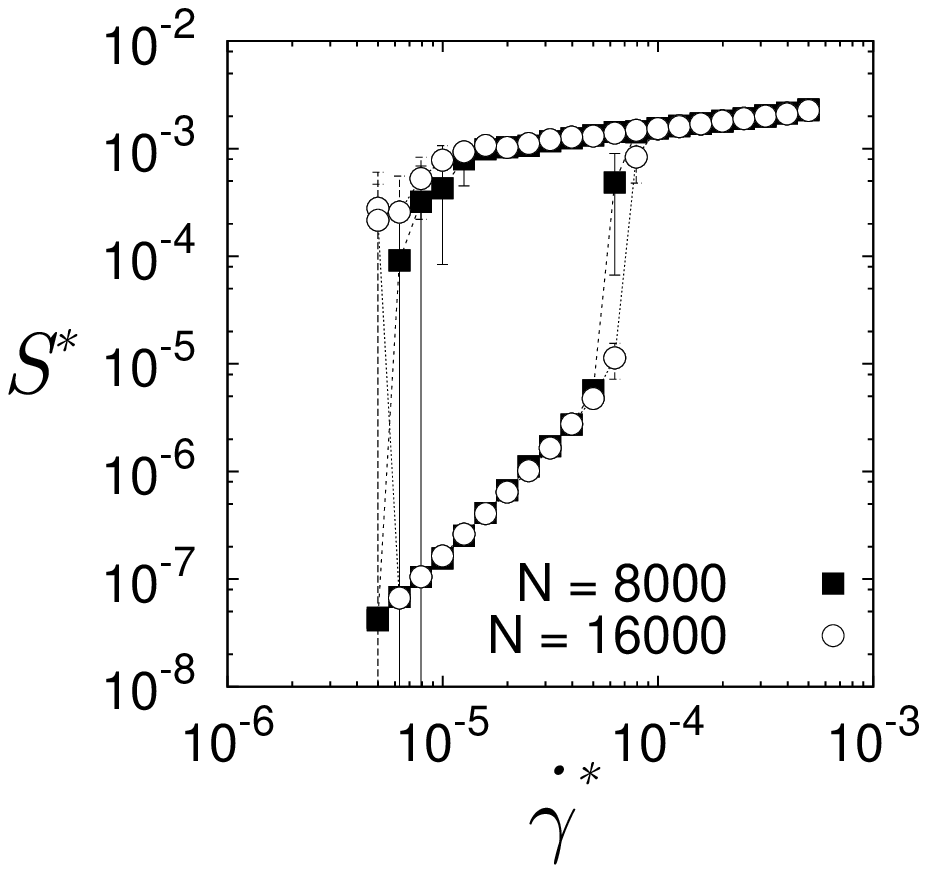} 
\caption{ 
  Scaled shear stress $S^* = S /(k^{(n)} \sigma_0^{-1})$ as a function of 
  scaled shear rate $\dot \gamma ^* = \dot \gamma \sqrt{k^{(n)}/m}$
  for $\phi = 0.795$, $\mu = 1.8$, and $a = 10^{0.1}$
  with $N=8000$ and $16000$.
  }
\label{rheology:size}
\end{center}
\end{figure}

Let us check how the hysteresis loop, 
as shown in Fig. \ref{rheology}, depends on the system size.
In Fig. \ref{rheology:size}, we examine
 the scaled shear rate $\dot \gamma ^* = \dot \gamma \sqrt{k^{(n)}/m}$ dependence of
the scaled shear stress $S^* = S /(k^{(n)} \sigma_0^{-1})$ 
for $\phi = 0.795$, $\mu = 1.8$,
and $a = 10^{0.1}$ with $N=8000$ and $16000$.
Although the critical  $\dot\gamma^*$ where the two branches appears slightly differ
between $N=8000$ and $16000$,
the shape of the hysteresis loop is almost unchanged.
Hence, we conclude that  the hysteresis loop still survives
even in the thermodynamic limit ($N\to \infty$).

\begin{figure}
\begin{center}
\includegraphics[width=8cm]{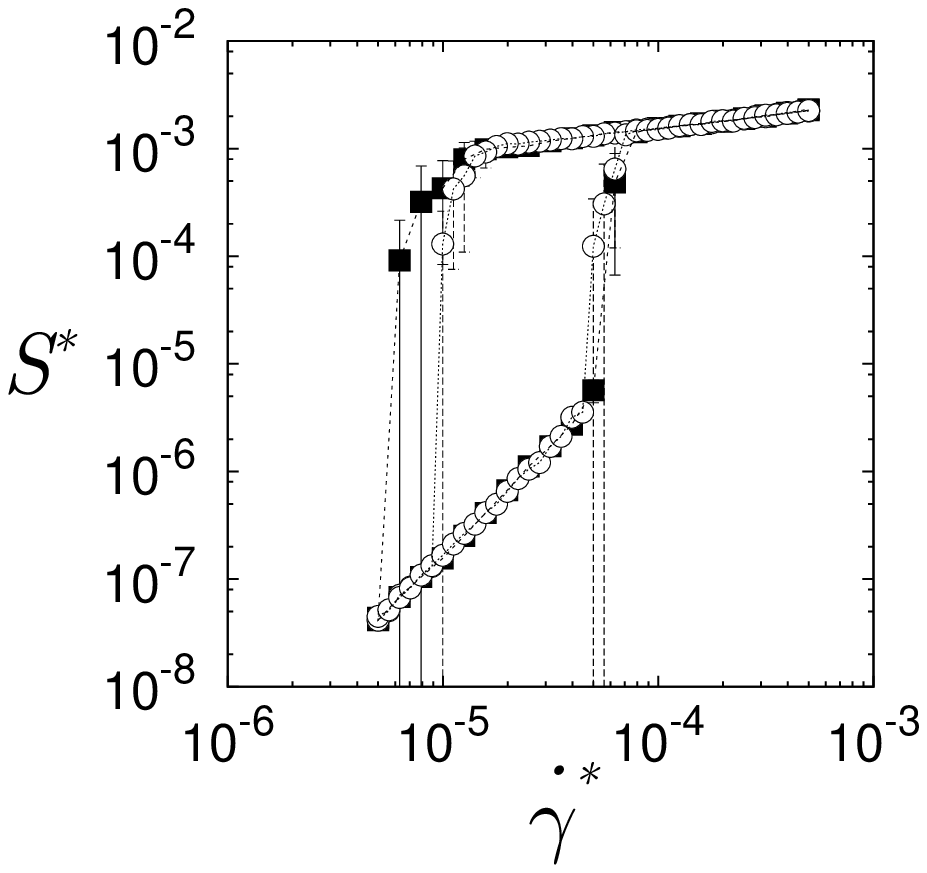} 
\caption{ 
  Scaled shear stress $S^* = S /(k^{(n)} \sigma_0)$ as a function of 
  scaled shear rate $\dot \gamma ^* = \dot \gamma \sqrt{k^{(n)}/m}$ 
  for $\phi=0.795$, $\mu = 1.8$ and $N=8000$ 
  with the shear rate step $a = 10^{0.1}$ characterized by the solid squares
  and $a = 10^{0.05}$ characterized by the open squares.
  }
\label{rheology:step}
\end{center}
\end{figure}

In Fig. \ref{rheology:step},
we compare the hysteresis loop of $S$
for $a = 10^{0.1}$
  with that for $a = 10^{0.05}$.
Owing to the slow change of the shear rate,
the transition from the {\it solid branch} to the {\it liquid branch}
for $a = 10^{0.05}$
takes place at a larger shear rate than that for $a = 10^{0.1}$.
However, the existence of the hysteresis loop and 
the values of $S$ in the {\it solid} and {\it liquid branches} are unchanged even when we use a smaller $a$.
Thus, qualitative behavior of the hysteresis loop is insensitive to the choice of the change rate of $\dot\gamma^*$.

Finally, we check the dependence of the hysteresis loops 
on the waiting time $\tau_{\rm w}$ for each shear rate.
In Fig. \ref{rheology:step},
we compare the hysteresis loop for 
$\tau_{\rm w} = 0.4 \dot \gamma^{-1}$ with that for 
$\tau_{\rm w} = 4.0 \dot \gamma^{-1}$ and
$\tau_{\rm w} = 8.0 \dot \gamma^{-1}$.
Although the area of the hysteresis loop for large waiting times is smaller than that of short waiting time,
the area is converged for $\tau_{\rm w}$ is larger than $4.0\dot\gamma^{-1}$. 
Thus, the hysteresis loop survives  
even when we slowly change the shear rate.

\begin{figure}
\begin{center}
\includegraphics[width=8cm]{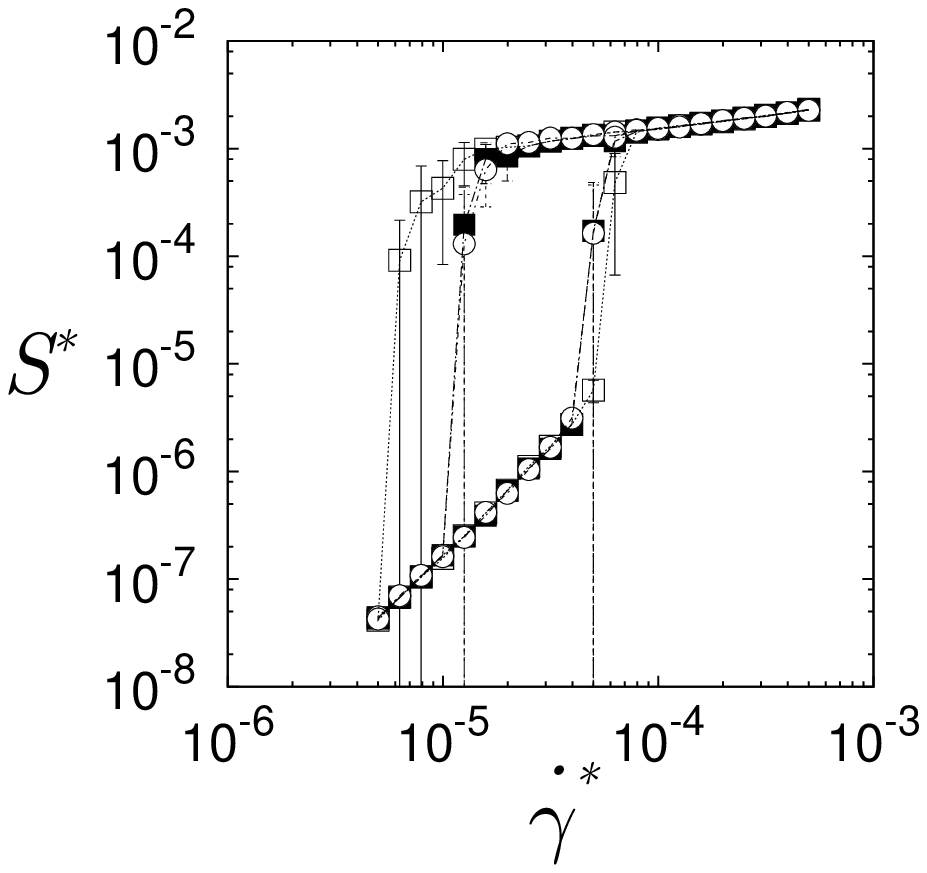} 
\caption{ 
  Scaled shear stress $S^* = S /(k^{(n)} \sigma_0)$ as a function of 
  scaled shear rate $\dot \gamma ^* = \dot \gamma \sqrt{k^{(n)}/m}$ 
  for $\phi=0.795$, $\mu = 1.8$, $a = 10^{0.1}$ and $N=8000$ 
    with the waiting time $\tau_{\rm w} = 0.4  \dot \gamma^{-1}$
    characterized by the open squares, $4.0 \dot \gamma^{-1}$
    characterized by the solid squares and $8.0 \dot \gamma^{-1}$
    characterized by the open circles.
  }
\label{rheology:time}
\end{center}
\end{figure}

\subsection{Critical densities and phase diagram}
\label{Critical:sec}

\begin{figure}
\begin{center}
\includegraphics[width=8cm]{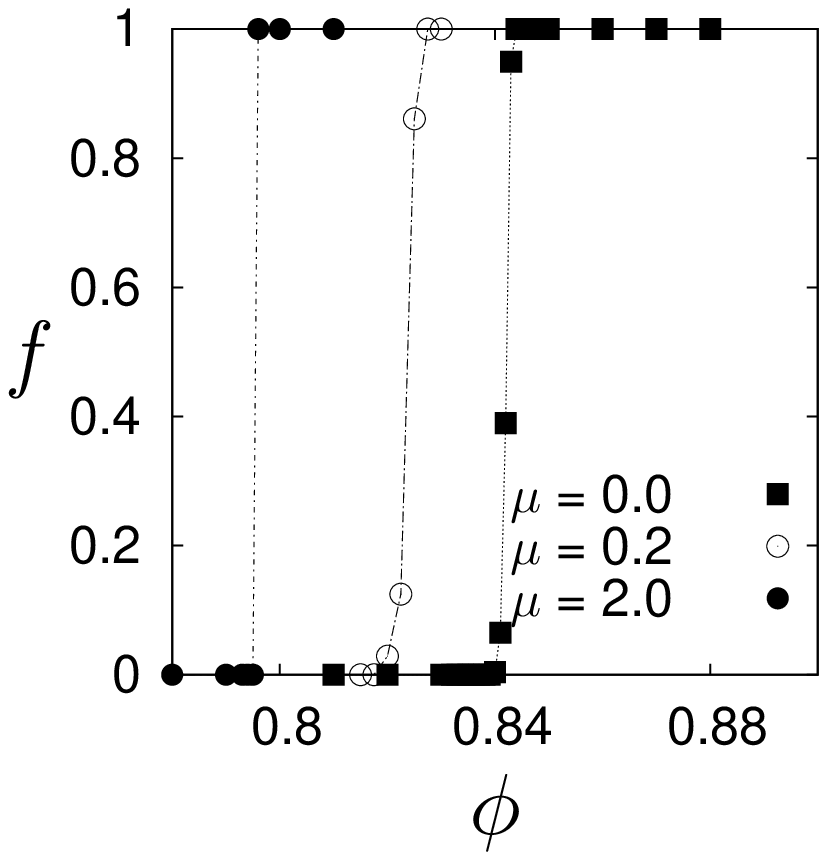} 
\caption{ 
  Jammed fraction $f$ as a function of $\phi$ for $N=4000$ with 
  $\mu = 0.0, 0.2$ and $2.0$.
  }
\label{f_fig}
\end{center}
\end{figure}

In this subsection, we demonstrate the existence of two critical densities.
First one is the transition density $\phi_C(\mu)$ at which
the shear stress $S$ in the low shear limit $\dot \gamma \to 0$ has 
a finite value.
In order to determine $\phi_C(\mu)$,
we introduce the jammed fraction $f$
obtained from the simulation using the QS method, 
where 
we introduce $f$ as a fraction of samples where $S$
is larger than a threshold value $S_{\rm th} = 10^{-7} k^{(n)}/\sigma_0$.
In Fig. \ref{f_fig}, we plot the jammed fraction as a function of the
area fraction $\phi$ for $\mu=0.0, 0.2$ and $2.0$.
The jammed fraction suddenly changes around $\phi=0.843, 0.825 $ and $0.795$
for  $\mu=0.0, 0.2$ and $2.0$, respectively.
Then, we can determine $\phi_C(\mu)$ as the area fraction at $f=0.5$.

\begin{figure*}
\begin{tabular}{cc}
\begin{minipage}{0.5\hsize}
\begin{center}
\includegraphics[width=8cm, bb=50 30 410 322 ]{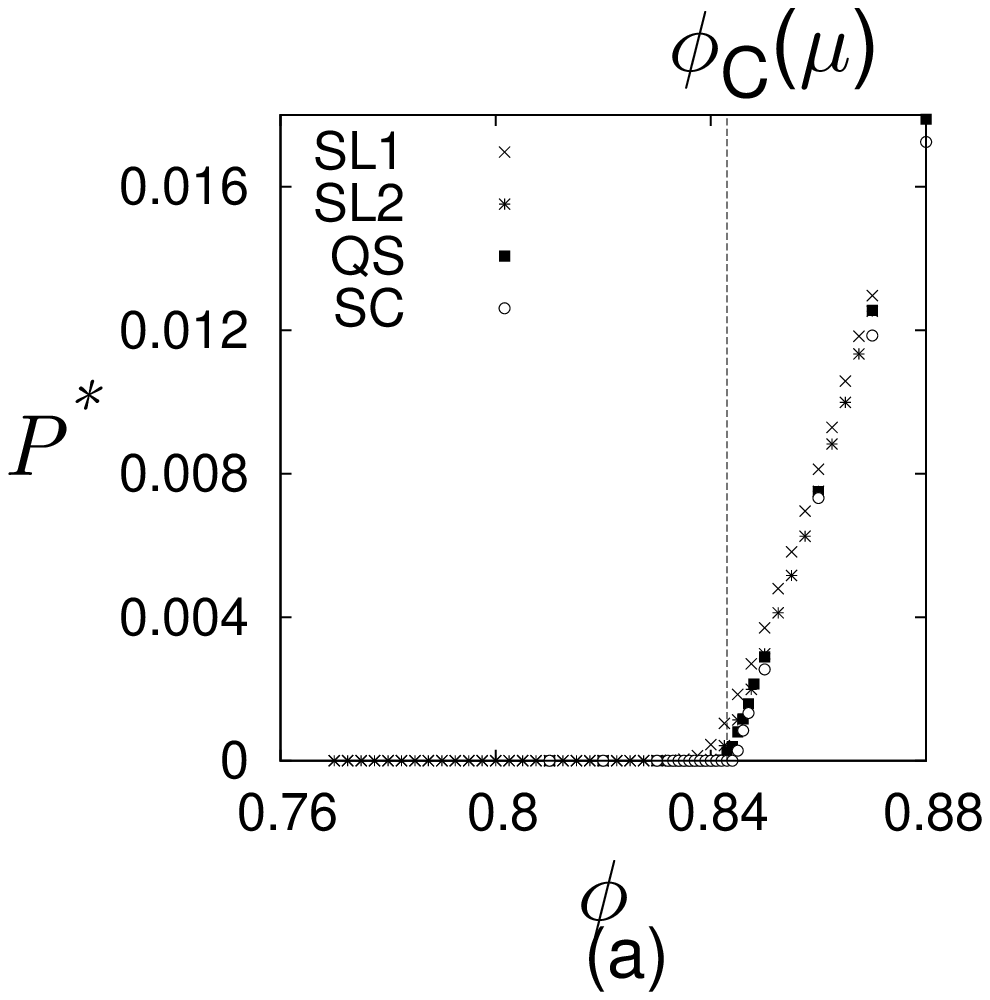} 
\end{center}
\end{minipage}
\begin{minipage}{0.5\hsize}
\begin{center}
\includegraphics[width=8cm, bb=50 30 410 322 ]{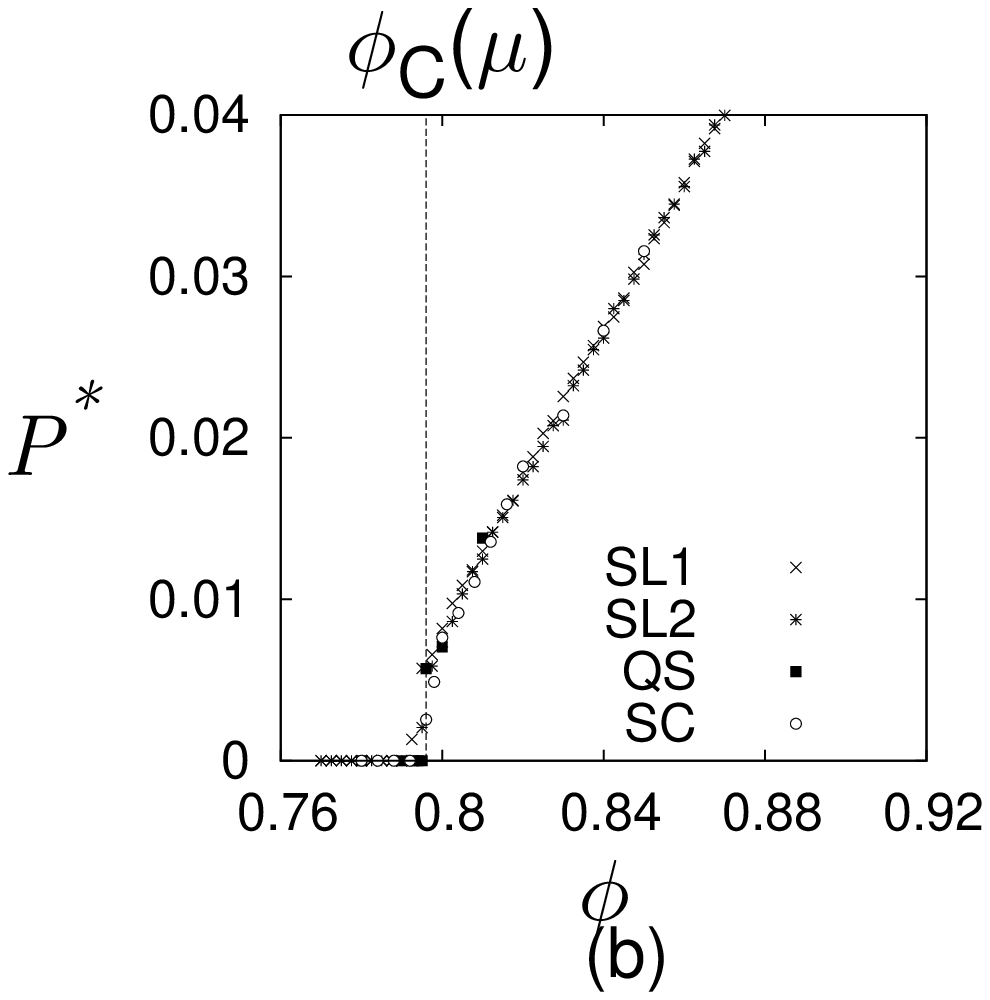} 
\end{center}
\end{minipage}
\end{tabular}
\begin{tabular}{cc}
\begin{minipage}{0.5\hsize}
\begin{center}
\includegraphics[width=8cm, bb=50 30 410 322 ]{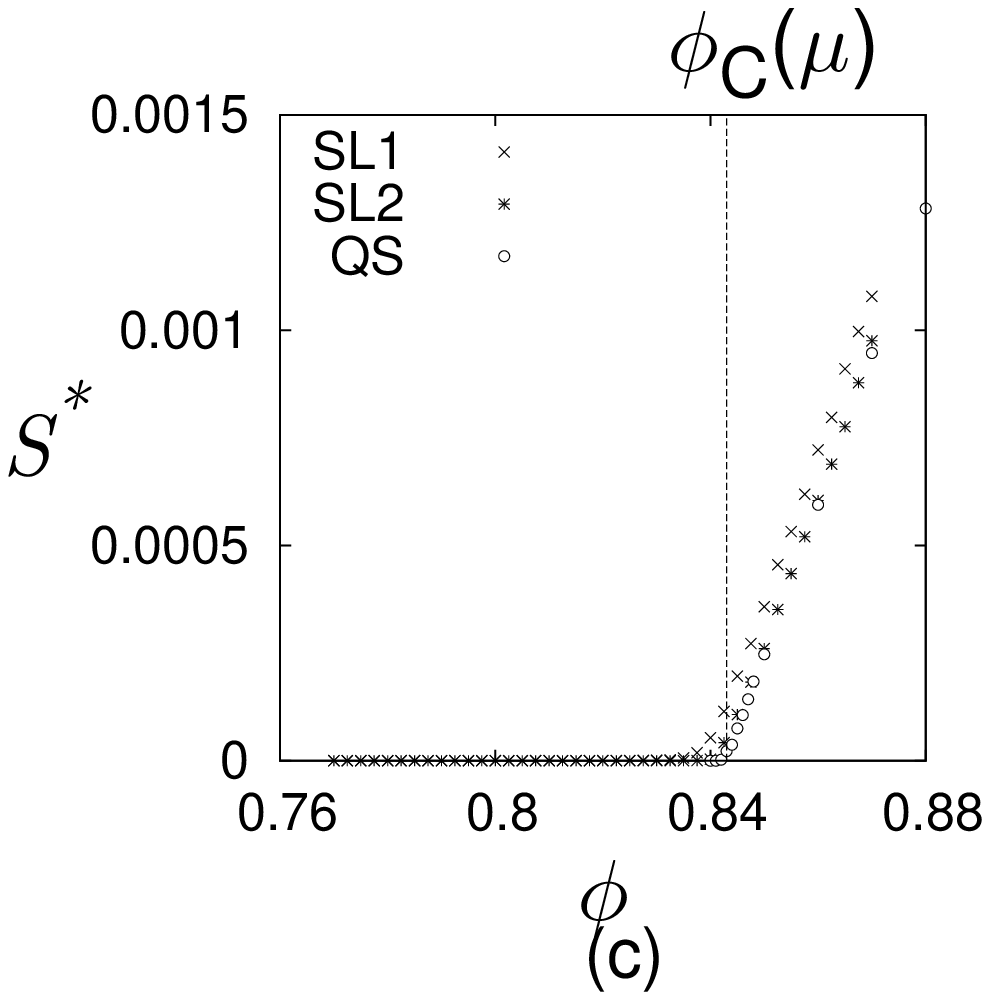} 
\end{center}
\end{minipage}
\begin{minipage}{0.5\hsize}
\begin{center}
\includegraphics[width=8cm, bb=50 30 410 322 ]{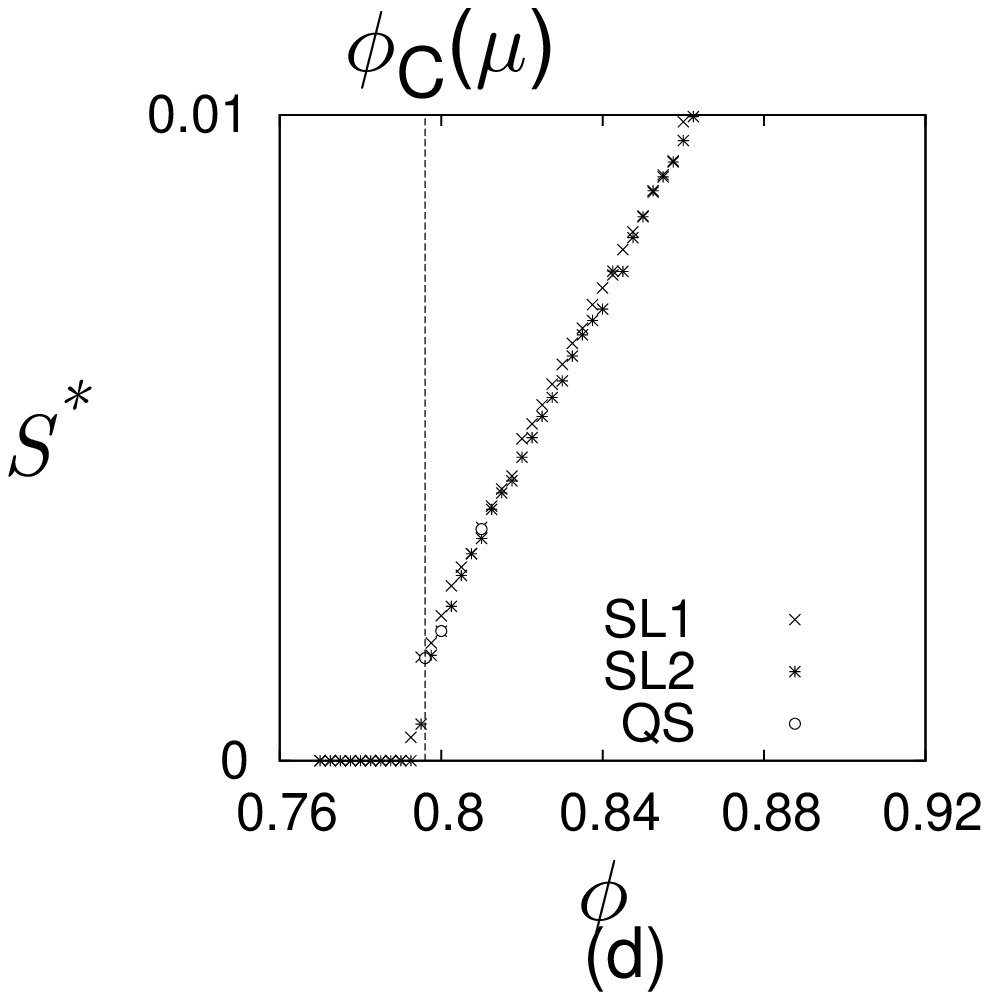} 
\end{center}
\end{minipage}
\end{tabular}
\caption{ 
  (a):The rescaled pressure $P^*$ with $P^* = P /(k^{(n)} \sigma_0^{-1})$
  as a function of $\phi$ for $N=8000$ by using
  the SL method with $\dot \gamma^* = 5 \times 10^{-5}$
  which we call ``SL1'', 
  the SL method with $\dot \gamma^* = 5 \times 10^{-6}$
  which we call  ``SL1'', 
  the QS, and the SC methods for $\mu=0.0$.
  (b):The rescaled pressure $P^*$ with $P^* = P /(k^{(n)} \sigma_0^{-1})$
  as a function of $\phi$ for $N=8000$ by using
  the SL method with $\dot \gamma^* = 5 \times 10^{-5}$
  which we call   ``SL1'', 
  the SL method with $\dot \gamma^* = 5 \times 10^{-6}$
  which we call   ``SL2'', 
  the QS, and the SC methods for $\mu=2.0$.
  (c):The rescaled stress $S^*$ with $S^* = S /(k^{(n)} \sigma_0^{-1})$
  as a function of $\phi$ for $N=8000$ by using
  the SL method with $\dot \gamma^* = 5 \times 10^{-5}$
  which we call   ``SL1'', 
  the SL method with $\dot \gamma^* = 5 \times 10^{-6}$
  which we call   ``SL2'', 
  and the SC methods for $\mu=0.0$.
  Note that we eliminate the data by SC. 
  (d):The rescaled stress $S^*$ with $S^* = S /(k^{(n)} \sigma_0^{-1})$
  as a function of $\phi$ for $N=8000$ by using
  the SL method with $\dot \gamma^* = 5 \times 10^{-5}$
  which we call   ``SL1'', 
  the SL method with $\dot \gamma^* = 5 \times 10^{-6}$
  which we call   ``SL2'', 
  and the SC methods for $\mu=2.0$.
}
\label{P_QS}
\end{figure*}

The second critical density is obtained from the pressure in the low shear limit.
In Fig. \ref{P_QS}, we plot the value of the pressure $P$ and the stress $S$
obtained from various methods SL, QS and SC
as a function of the area fraction $\phi$ for $\mu=0.0$ and $2.0$.
As shown in Fig. \ref{P_QS}, $P$ and $S$ converge to the value of the QS method
in the low shear limit of the SL method.
For the SC method, if we take small enough $S_0$, the value of $P$
is almost equal to that of the QS method as shown in Fig. \ref{P_QS},
while $S$ becomes almost zero.
Here, we should note that the pressure $P$ and the stress $S$
continuously increase
from the transition point $\phi_C(\mu)$ for $\mu = 0.0$,
while $P$ and $S$ in the low shear limit 
discontinuously change at $\phi_C(\mu)$ for $\mu=2.0$.
This discontinuous change should 
be related to the emergence of a hysteresis loop. 
Therefore, $\phi_C(\mu)$ can be regarded as a true critical area fraction for the jamming transition.

\begin{figure*}[htbp]
\begin{tabular}{cc}
\begin{minipage}{0.5\hsize}
\begin{center}
\includegraphics[width=8cm, bb=50 20 410 302 ]{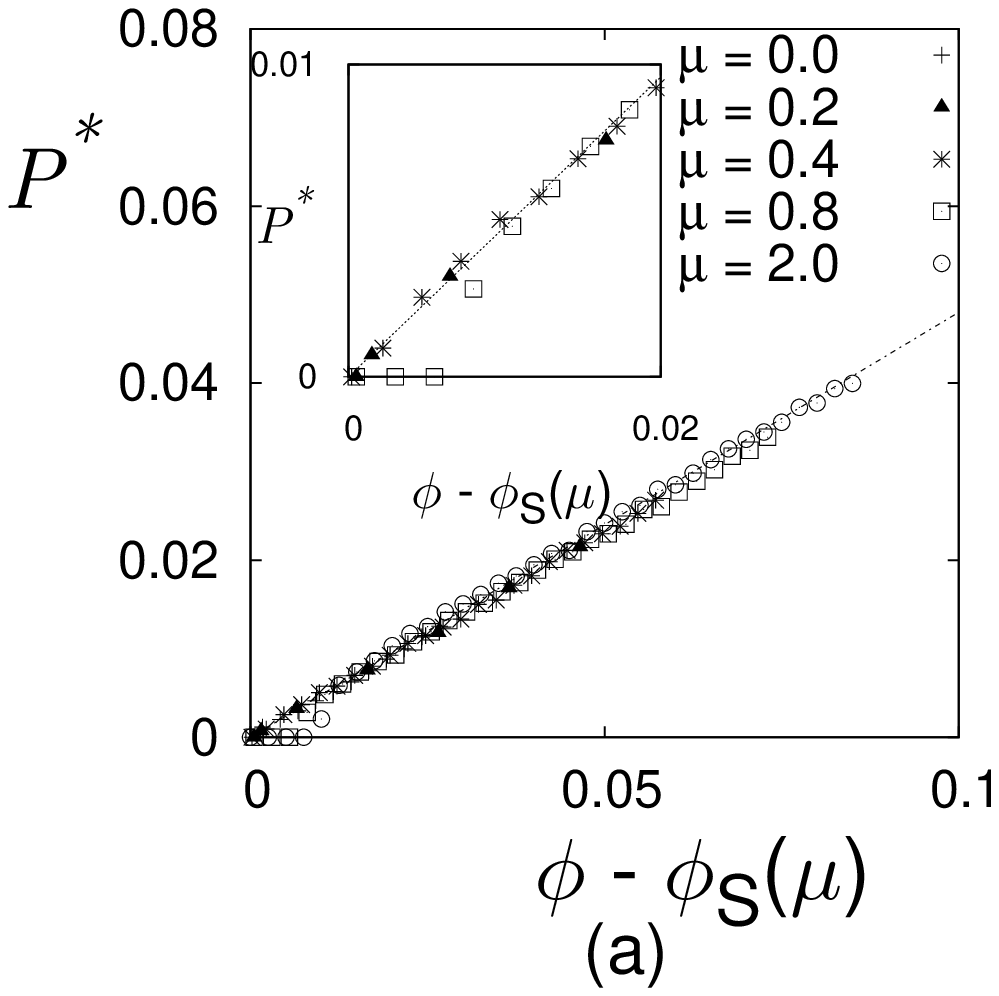} 
\end{center}
\end{minipage}
\begin{minipage}{0.5\hsize}
\begin{center}
\includegraphics[width=8cm, bb=50 20 410 302 ]{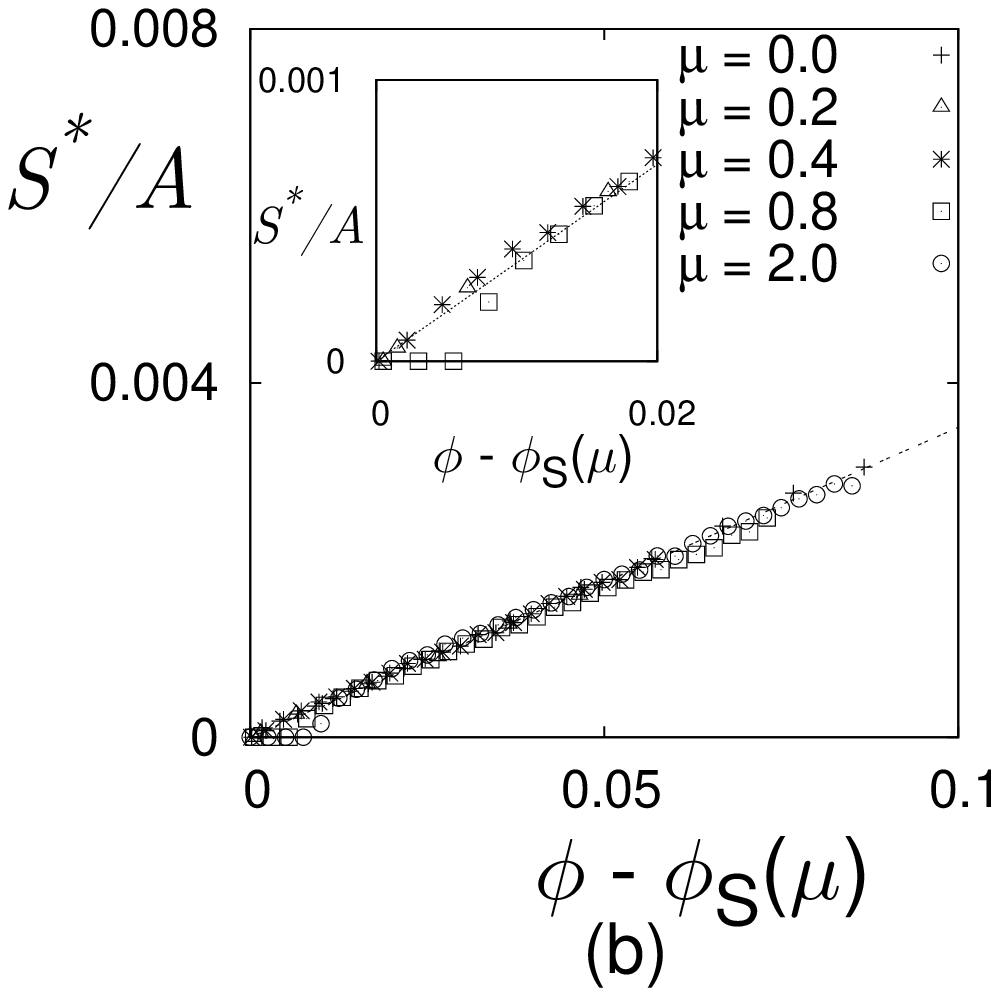} 
\end{center}
\end{minipage}
\end{tabular}
\caption{ 
  (a): The rescaled pressure $P^*$ with $P^* = P /(k^{(n)} \sigma_0^{-1})$
  as a function of $\phi - \phi_S(\mu)$ for $N=8000$
  with $\dot \gamma = 5.0 \times 10^{-6}$.
  Inset shows the rescaled pressure $P^*$ 
  near $\phi_S(\mu) - \phi = 0$ for $\mu=0.2, 0.4, 0.8$.
   (b) : 
 $S^* / A$ with $S^* = S /(k^{(n)} \sigma_0^{-1})$
  for $\dot \gamma = 5.0 \times 10^{-6}\sqrt{k^{(n)}/m}$ and $N=8000$
  as a function of $\phi - \phi_S(\mu)$ in the {\it solid branch},
  where $A$ is a constant that depends only on $\mu$ i.e., 
$A=1.0, 0.4, 0.3, 0.3$, and $0.3$
for $\mu=0.0, 0.2, 0.4, 0.8$, and $2.0$, respectively. 
  Inset shows $S^* / A$ near $\phi_S(\mu) - \phi = 0$ for $\mu=0.2, 0.4, 0.8$.  
  }
\label{P_scale_phi}
\end{figure*}

On the other hand, we find that a critical scaling is satisfied
if we introduce another critical point $\phi_S(\mu)$ as
\begin{equation}
P(\phi, \mu) = \Pi_P(\phi - \phi_S(\mu)), \quad S(\phi, \mu) / A(\mu) = \Pi_S(\phi - \phi_S(\mu))
\label{P_QS_scale}
\end{equation}
for $\mu_S(\mu)\le \phi_C(\mu)$, where the denominator $A(\mu)$ in the
second equation depends only on $\mu$. Figure \ref{P_scale_phi} clearly verifies the
validity of the scaling relation \eqref{P_QS_scale}, though the data for
$\phi<\phi_C(\mu)$ are not involved. Here, we choose
$\phi_S(0)=\phi_C(0)$ for $\mu=0$, while we choose $\phi_S(\mu$ from
the collapse of the data onto the universal scaling curve for $\mu>0$
(Fig. \ref{P_scale_phi}).
Our result also suggests that $\Pi_P(x)$ and $\Pi_S(x)$ are linear functions of $x$ for $x>0$.
In the inset of Fig. \ref{P_scale_phi},
we enlarge the region near $\phi_S(\mu) - \phi = 0$,
which indicates that $P$ and $S$ discontinuously change for $\mu\ge 0.8$, while they seems to have continuous changes for $\mu\le 0.4$.
Here, we should note that $\phi_S(\mu)$ is a fictitious critical fraction, at least for $\mu\ge 0.8$, 
but is important in characterizing scaling laws  for the rheology of the frictional
granular materials  in the next subsection.

\begin{figure}
\begin{center}
\includegraphics[width=8cm]{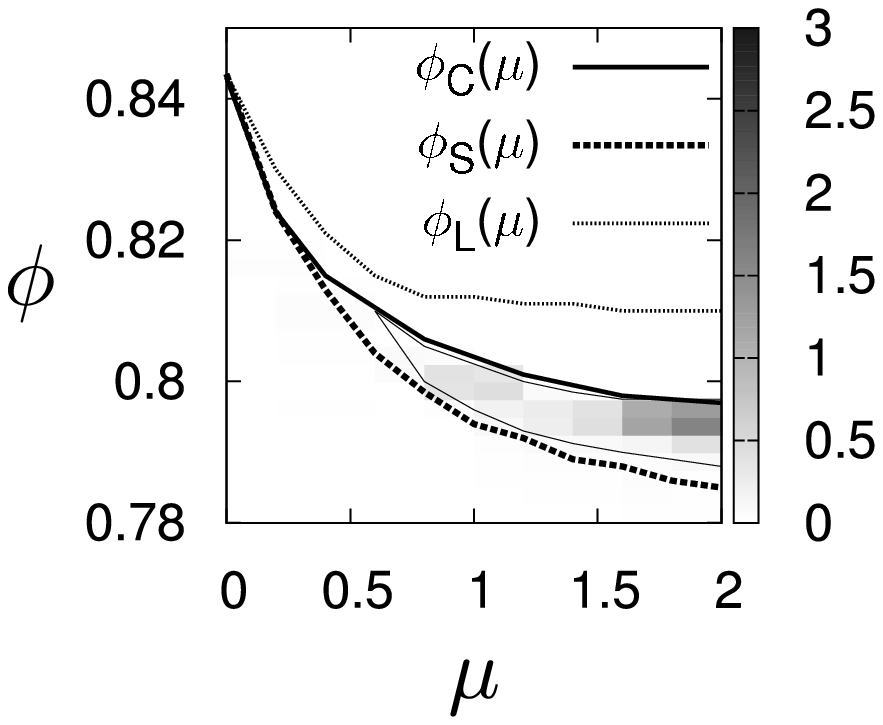}
\caption{ 
  The critical fractions $\phi_C(\mu)$, $\phi_S(\mu)$, and $\phi_L(\mu)$
  as a function of $\mu$.
  The amount $\alpha$ of the hysteresis-dependence in $\mu-\phi$ plane
  is plotted with a gray scale.
  The region of the hysteresis loop,
  which is defined by the region where $\alpha>0.1$,
  is enclosed by thin solid lines.
  }
\label{phase_fig}
\end{center}
\end{figure}

In Fig. \ref{phase_fig}, we show the dependence of the critical fractions
$\phi_C(\mu)$ and $\phi_S(\mu)$ on the friction coefficient $\mu$.
Both critical fractions decrease as  $\mu$ increases.
This result is similar to the previous ones for 
three-dimensional frictional spheres 
 \cite{Silbert02, Zhang, Shudyak,Somfai,Hecke,Henkes,Silbert10}.
As shown in Fig. \ref{phase_fig},
the difference between $\phi_C(\mu)$ and $\phi_S(\mu)$ is not visible for $\mu<0.4$,
but it becomes distinct as $\phi_C(\mu) > \phi_S(\mu)$ for $\mu>0.4$.
This discrepancy between $\phi_C(\mu)$ and $\phi_S(\mu)$
is related with the discontinuous change of $P$ shown in
Figs. \ref{P_QS} and \ref{P_scale_phi}.
Figure \ref{phase_fig} also exhibits  
the region where the hysteresis appears
for $N=8000$ and $a = 10^{0.1}$.
Here, let us introduce  $\alpha$ to represent 
the area of the hysteresis loop shown in Fig. \ref{rheology}(a)
\begin{equation}
\alpha \equiv \int_{G_1}^{G_2} dG \ \left \{ 
  \log_{10} S_+(\dot \gamma) - \log_{10} S_-(\dot \gamma)
  \right \},
\end{equation}
where $S_+(\dot\gamma)$ and $S_-(\dot\gamma)$ respectively represent the shear stresses in the {\it decreasing}
and the {\it increasing process}
for $G \equiv \log \dot \gamma$, 
$G_1 = \log_{10} \left ( \dot \gamma_0 a^{-N_s} \right )$,
and $G_2 = \log_{10} \left ( \dot \gamma_0 \right )$.
In Fig. \ref{phase_fig},
we plot $\alpha$ in the $\mu$-$\phi$ plane
with a gray scale.
The hysteresis only appears in a restricted density region for each 
values of $\mu$,
and the range of the region becomes wider as the friction coefficient $\mu$
increases.
In Fig. \ref{phase_fig}, we also enclose 
the region of the hysteresis loop by thin solid lines,
where
$\alpha$ is larger than a threshold value $\alpha_{\rm th}=0.1$. 
The boundary lines are almost identical to $\phi_C(\mu)$ and $\phi_S(\mu)$ and 
the hysteretic region lies between $\phi_S(\mu)$ and $\phi_C(\mu)$, 
although a discrepancy from the critical densities exist.
We suppose that the boundary lines coincide with $\phi_C(\mu)$
and $\phi_S(\mu)$ if we take a limit $\alpha_{\rm th} \to 0$.
However, if we use smaller $\alpha_{\rm th}$ in the present numerical data,
we could not draw the clear lines due to sampling error.
In this figure, the hysteresis loop is only verified in the region $\mu > 0.5$, 
though we cannot exclude the possibility of the existence of a small hysteresis loop for smaller $\mu$.
We note that Fig. \ref{phase_fig} involves the other fictitious critical fraction $\phi_L(\mu)$,
which will be introduced in the next subsection.

\begin{figure}
\begin{center}
\includegraphics[width=9cm]{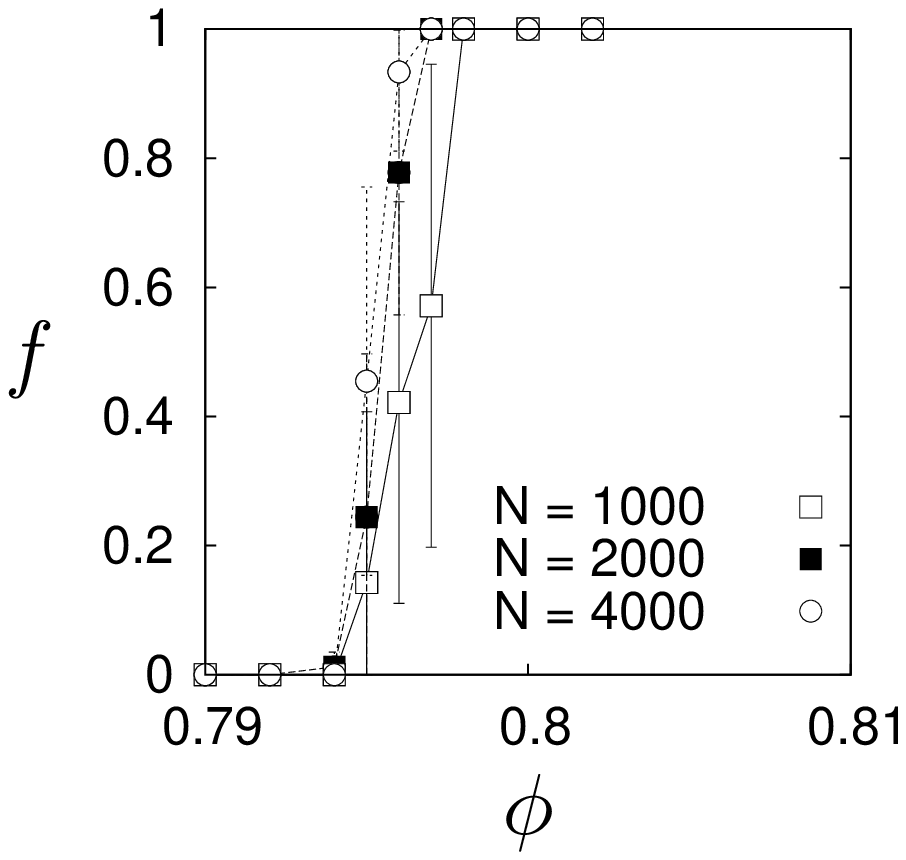} 
\caption{ 
  Jammed fraction $f$ as a function of $\phi$ for $\mu = 2.0$ with 
  $N = 1000, 2000$ and $4000$
obtained from the simulation using the QS method with
$\Delta \gamma = 10^{-6}$ and $E_{\rm th} = 10^{-7} k^{(n)}\sigma_0^2$.
 }
\label{f_size}
\end{center}
\end{figure}

Let us check the finite size effect of the critical fraction $\phi_C(\mu)$.
In Fig. \ref{f_size}, we examine
the jammed fraction $f$ as a function of $\phi$ for $\mu = 2.0$ with 
$N = 1000, 2000$ and $4000$.
Fig. \ref{f_size} indicates the existence of a finite size effect,
where $\phi_C(\mu)$ lies between $0.794$ and $0.798$.
However, this finite size effect does not affect the qualitative behavior
of the phase diagram in Fig. \ref{phase_fig}
because the difference between  $\phi = 0.794$ and $0.798$
is so small that it cannot be distinguished  in Fig. \ref{phase_fig}.

\subsection{Critical scaling laws}
\label{Scaling:sec}

\begin{figure}
\begin{center}
\includegraphics[width=8cm, bb=50 30 410 302 ]{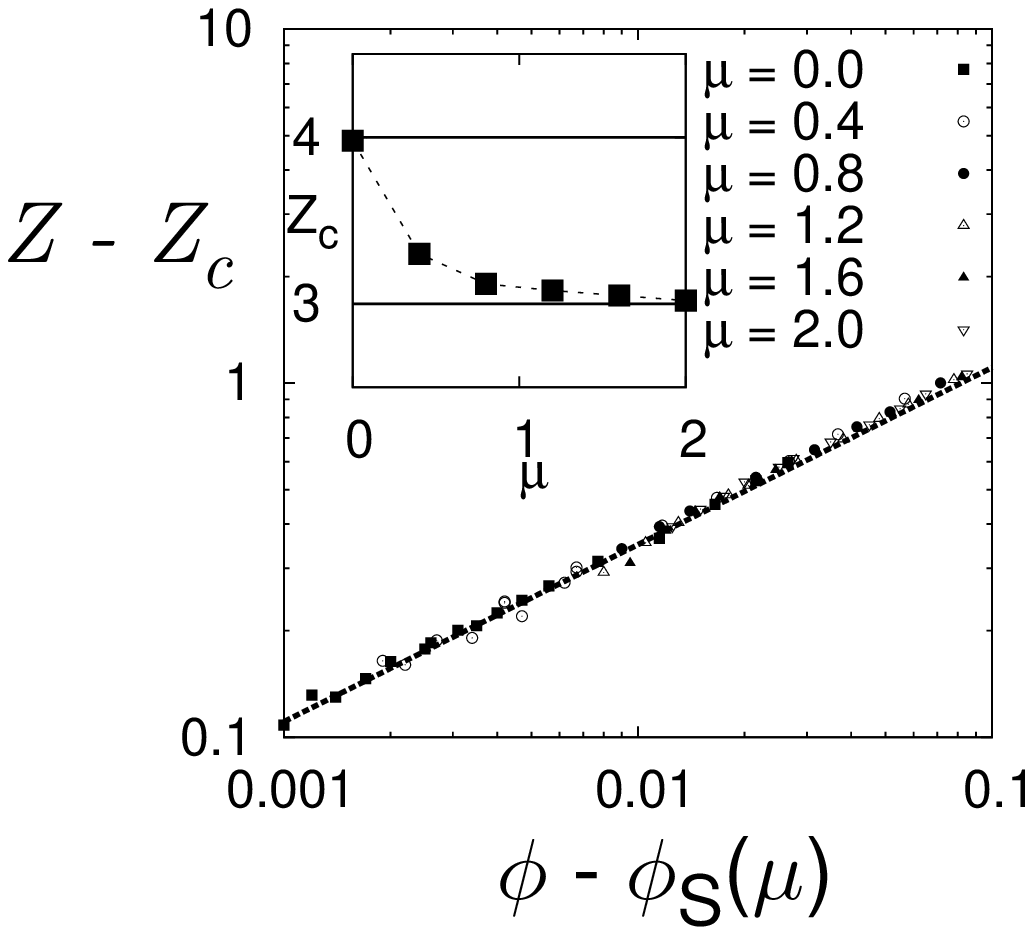} 
\caption{ 
  $Z - Z_c(\mu)$ 
  for $\dot \gamma = 5.0 \times 10^{-6}\sqrt{k^{(n)}/m}$ 
  as a function of $\phi - \phi_S(\mu)$ in the {\it solid branch}.
  Inset shows $Z_c(\mu)$ as a function of the friction coefficient $\mu$.
}
\label{jam_scale}
\end{center}
\end{figure}

In a frictionless system with the linear spring repulsion,
the shear stress $S$, the pressure $P$ and the coordination number $Z$
for $\phi > \phi_J$
satisfy 
 $S \propto \phi - \phi_J$, $P \propto \phi - \phi_J$,
and $Z - Z_c \propto (\phi - \phi_J)^{1/2}$, where $Z_c=4$
is the coordination number for the isostatic state of frictionless
particles \cite{Otsuki08,Otsuki09}.
Remarkably, these scaling laws are still valid
even for frictional systems
if we use $\phi_S(\mu)$ defined in Sec. \ref{Critical:sec} as
\begin{equation}
S \sim \phi - \phi_S(\mu), \ P \sim \phi - \phi_S(\mu), \
  Z - Z_c(\mu) \sim \sqrt{\phi - \phi_S(\mu)}.
 \label{jam:eq}
\end{equation}
Figure \ref{P_scale_phi} supports the scaling relations \eqref{jam:eq}
for the pressure $P$ and the shear stress $S$.
Here, it should be noted that
the scaling relation $P \sim (\phi - \phi_J)^{1.08}$,
which is reported in Ref. \cite{Olsson10}, might be better than
Eq. \eqref{jam:eq} for $\mu=0.0$,
but the data for $\mu>0$ suggests that the linear relationship between $P$ and $\phi-\phi_S(\mu)$ might be better than
$P \sim (\phi - \phi_C(\mu))^{1.08}$.
In Fig. \ref{jam_scale}, we plot  $Z$
as functions of $\phi - \phi_S(\mu)$,
which also supports the scaling relation \eqref{jam:eq} for $Z$.
The inset in Fig. \ref{jam_scale} shows the $\mu$-dependence of 
$Z_c$, which approaches the critical value ($Z_c = 3$) for isostatic
frictional grains.
This behavior of $Z_c(\mu)$ is similar to the previous
result obtained for the static granular packing problem
\cite{Silbert02, Zhang, Shudyak,Hecke,Silbert10}.
It should be noted that only the data in the {\it solid branch}
satisfy the scaling relation Eq. \eqref{jam:eq}.

\begin{figure*}
\begin{center}
\includegraphics[width=0.8 \linewidth]{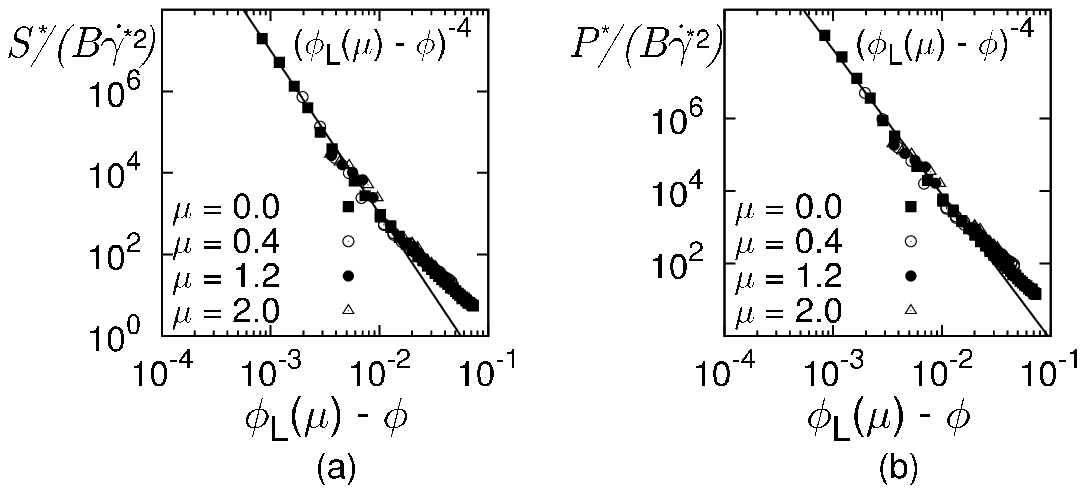}
\caption{ 
  (a): $S^* / (B \dot \gamma^{*2})$ 
  with $S^* = S /(k^{(n)} \sigma_0^{-1})$ and $\dot \gamma^*  = \dot \gamma 
 \sqrt{k^{(n)}/m}$ 
  for $\dot \gamma^* = 2.0 \times 10^{-7}$ and $N=30000$ 
  as a function of $\phi_L(\mu) - \phi$ in the {\it liquid branch},
  where $B$ is a constant that depends only on $\mu$ i.e., 
$B=1.0, 0.5, 0.25$, and $0.25$,
for $\mu=0.0, 0.4, 1.2$, and $2.0$, respectively. \\
  (b): $P^* / (B \dot \gamma^{*2})$ 
  with $P^* = S /(k^{(n)} \sigma_0^{-1})$ and $\dot \gamma^*  = \dot \gamma 
 \sqrt{k^{(n)}/m}$ 
  for $\dot \gamma^* = 2.0 \times 10^{-7}$  and $N=30000$ 
  as a function of $\phi_L(\mu) - \phi$ in the {\it liquid branch},
  where $B$ is a constant that depends only on $\mu$ i.e., 
$B=1.0, 0.9, 0.25$, and $0.25$,
for $\mu=0.0, 0.4, 1.2$, and $2.0$, respectively.
}
\label{unjam_scale}
\end{center}
\end{figure*}

For $\phi < \phi_J$ in a frictionless system 
it is known that Bagnold's scaling holds, and
$S$ and $P$ satisfy
$S \propto  \dot \gamma^2(\phi_J - \phi)^{-4}$ and 
$P \propto  \dot \gamma^2(\phi_J - \phi)^{-4}$.
Even in the frictional system,
$S$ and $P$ satisfies Bagnold's scaling 
$S \propto \dot \gamma^2$ and $S \propto \dot \gamma^2$
in the {\it liquid branch} as shown in Fig. \ref{rheology},
but any scaling relation
using $\phi_C(\mu)$ and $\phi_S(\mu)$ cannot be applied.
However, by introducing another fictitious critical fraction $\phi_L(\mu)$
shown in Fig. \ref{phase_fig},
$S$ and $P$ satisfy the scaling relations
\begin{equation}
S \sim \dot \gamma^2 \{ \phi_L(\mu) - \phi \}^{-4} \quad
\mbox{and} \quad
P \sim \dot \gamma^2 \{ \phi_L(\mu) - \phi \}^{-4}
\label{liquid:eq}
\end{equation}
in the {\it liquid branch}.
In Fig. \ref{unjam_scale}, 
we plot $S/\dot \gamma^2$ and $P/\dot \gamma^2$ in the {\it liquid branch},
where Eq. \eqref{liquid:eq} is satisfied.
However, $\phi_L(\mu)$ is just a fitting parameter for  Eq. \eqref{liquid:eq},
and could not be estimated from an independent protocol.
In this sense, Eq. (\ref{liquid:eq}) might be superficial.

We should note that
 the scalings of $Z$, $P$ and $S$ in the jammed phase
of the static granular packing problem
are not affected by the introduction of the friction
\cite{Somfai,Hecke,Silbert10}.
 Equations \eqref{jam:eq}  and \eqref{liquid:eq}
extend the validity of such an idea even for a system 
in the sheared granular systems.

It should be noted that the range of the excess density for the scaling of 
the {\it liquid branch} in Fig. \ref{unjam_scale}
is at most one decade because the shear rate should be smaller
to obtain the {\it liquid branch} near the critical point.
However, the validity of the 
scaling in Eq. \eqref{liquid:eq} for the frictionless system in the hard core limit
has been already verified in Ref.
\cite{Otsuki10} in details.

\section{Discussion and conclusion}
\label{Discussion:sec}

Let us compare our results with those of the previous studies
for the static granular packing problem.
In Refs. \cite{Silbert02, Zhang, Shudyak,Somfai,Hecke,Henkes,Silbert10},
the $\mu$-dependence of the scaling laws and the critical point
are studied,
where the critical fraction decreases as the friction coefficient $\mu$ increases.
These results are consistent with our results for the {\it solid} branch.
However, the results corresponding to those for the {\it liquid} branch 
in Figs. \ref{phase_fig} and \ref{unjam_scale} 
have not been reported in any previous studies.
Thus, we stress that the scaling laws with the aid 
of the $\mu$-dependence of $\phi_L(\mu)$ are our new findings.

Moreover, similar phase diagrams to Fig. \ref{phase_fig}
are proposed in Refs. \cite{Hecke,Somfai_v1}.
However, their results suggest that critical densities continuously exist between two critical values,
while our results indicate that the critical densities are at most three.
This is the main difference between
the phase diagram in Refs. \cite{Hecke,Somfai_v1} and ours.

In Ref. \cite{Ciamarra09},
it has been already reported that 
the critical fraction splits for 
frictional granular particles driven by a constant force
from the boundary wall.
There are three critical fractions which are denoted by $\phi_{J_1}$,
$\phi_{J_2}$ and $\phi_{J_3}$.
Here, $\phi_{J1}$ is the critical fraction used to characterize the appearance of the jamming transition,
$\phi_{J2}$ is that at which the viscosity $\eta$ diverges,
and $\phi_{J3}$ is that from where the pressure  increases.
Thus, we expect that these critical fractions $\phi_{J1}$,  $\phi_{J2}$,
and $\phi_{J3}$ respectively
correspond to $\phi_C(\mu)$, $\phi_L(\mu)$, and $\phi_S(\mu)$ in our system.
However, these critical fractions satisfy $\phi_{J1} <\phi_{J2} <\phi_{J3}$,
in contrast to the relation $\phi_{S}(\mu) <\phi_{C}(\mu) <\phi_{L}(\mu)$
We plan to clarify the origin of 
these differences in the future work.

In Ref. \cite{Somfai}, the scaling relations are 
plotted as a function of the distance from
isostaticity $Z - (D+1)$ with the dimension $D$ of the system,
where $D+1$ is the coordination number of the isostatic state.
However, we adopt the scaling relation as a function of the distance
from the critical density $\phi - \phi_S(\mu)$ in Fig. \ref{jam_scale}.
This is because $Z$ is not a control parameter and the scaling using
 $Z - (D+1)$ cannot be used for the {\it liquid branch},
where $Z$ tends to zero for sufficiently hard grains.

In previous studies for sheared frictionless materials,
the different scaling from Eq. \eqref{jam:eq} are proposed
\cite{Olsson,Hatano08,Tighe,Hatano10}.
In particular, $S \sim (\phi - \phi_J)^{3/2}$ 
for the jammed granular phase is used
in Ref. \cite{Tighe,Hatano10}.
It should be noted that the difference between ours and Hatano's \cite{Hatano10} exists only in this relationship,
but there are several differences between Hatano's \cite{Hatano10} and the exponents by Tighe et al. \cite{Tighe}.
We also note that there are no similarities between Tighe et al. \cite{Tighe} and Olsson and Title \cite{Olsson}.
Fortunately, our scaling relations $S\propto \phi-\phi_S(\mu)$ and $P\propto \phi-\phi_S(\mu)$  in the jammed region is supported by the simulation presented in Figs. \ref{P_scale_phi} and 10.
Moreover, the exponent is close to the exponent 1.08 obtained by Olsson and Title \cite{Olsson10},
although they claim that 1.08 is an evidence for non-linear behavior.
It should be noted that the scaling relations in Refs.
\cite{Otsuki08,Otsuki09,Otsuki10} are obtained for
granular particles with inertia effect, and the scaling relations are held in hard-core and elastic limit.
Therefore, our scaling exponents are  not necessary to be identical to those
  in Ref. \cite{Olsson,Tighe} for sheared foams without inertia effect.
Indeed, the derivation of our exponents is due  to the propagation of phonon under the isostatic condition \cite{Wyart05}.
On the other hand, Hatano\cite{Hatano10} indicated the estimation of the critical exponent depends on
the choice of the value of the critical density $\phi_J$ and he suggested $P \propto (\phi-\phi_J)^{1.5}$ from his simulation.
However, as shown in Sec. \ref{Scaling:sec},
if we chose the value of $\phi_S(\mu)$ which is estimated from the data
under the three different methods,
our scaling exponent seems to be better than his \cite{Hatano10}.
Here, we should note that the crossover from $P \sim (\phi - \phi_J)$
to $P \sim (\phi - \phi_J)^{3/2}$ might be understood from
the deviation from the critical point as $P \sim Z(\phi)(\phi - \phi_J)$
with $Z(\phi) - Z_c \propto (\phi - \phi_J)^{1/2}$.
We will have to resolve the contradiction between ours and his in near future.
We finally note that the data reported in Ref. \cite{Hecke} also suggests that $P\propto (\phi-\phi_J)^{3/2}$,
but his data is obtained from the Hertzian contact model, where the exponent 3/2 is equivalent to ours \cite{Otsuki08,Otsuki09}.

\begin{figure}
\begin{center}
\includegraphics[width=8cm, bb=50 30 410 302 ]{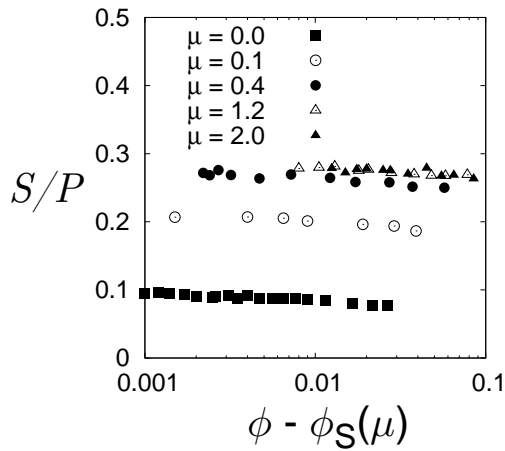} 
\caption{ 
  $S/P$ in the {\it solid branch} as a function of $\phi - \phi_S$
  for $\dot \gamma = 5.0 \times 10^{-6}\sqrt{k^{(n)}/m}$ and $N=8000$. 
  }
\label{S_P_jam}
\end{center}
\end{figure}

\begin{figure}
\begin{center}
\includegraphics[width=8cm, bb=50 30 410 302 ]{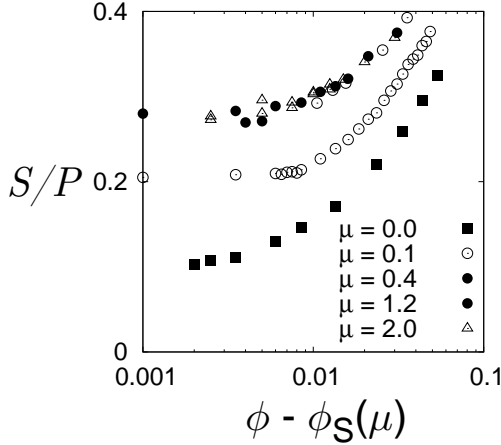} 
\caption{ 
  $S/P$ in the {\it liquid branch}
  as a function of $\phi_L(\mu) - \phi$
  for $\dot \gamma = 2.0 \times 10^{-7}\sqrt{k^{(n)}/m}$ and $N=30000$.
  }
\label{S_P_unjam}
\end{center}
\end{figure}

\begin{figure}
\begin{center}
\includegraphics[width=8cm]{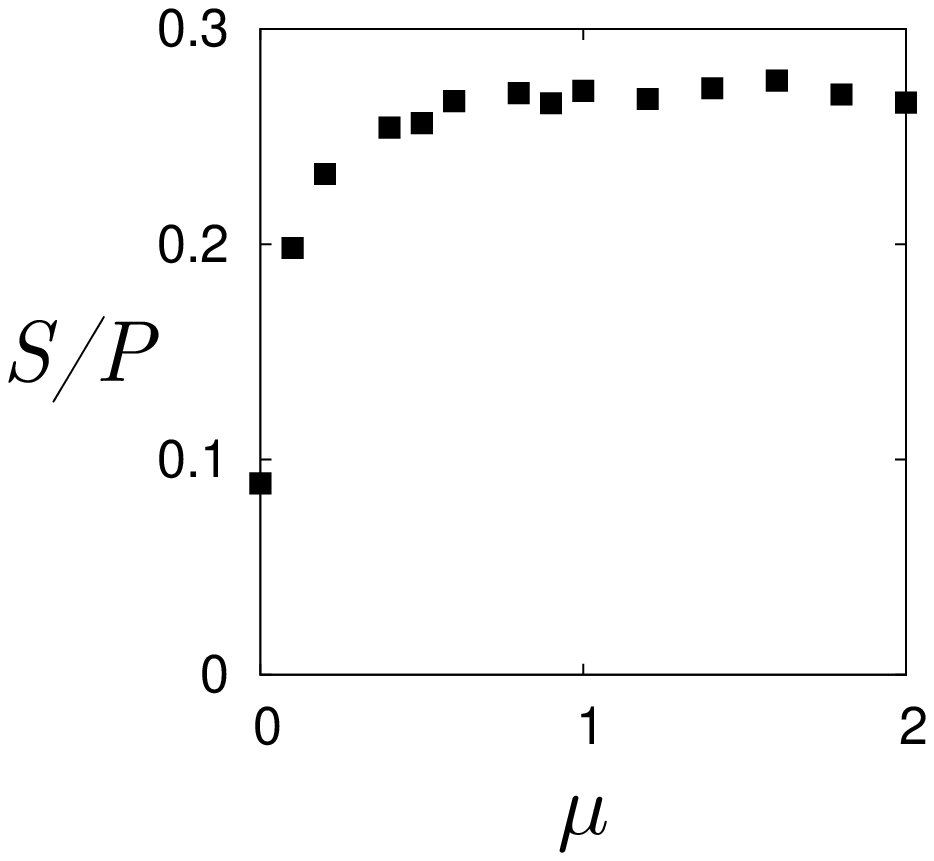} 
\caption{ 
  $S/P$ in the {\it solid branch} as a function of $\mu$
  for $\phi = 0.850$, $\dot \gamma = 5.0 \times 10^{-6}\sqrt{k^{(n)}/m}$,
    and  $N=8000$.
  }
\label{mu}
\end{center}
\end{figure}

Here, we should discuss the {\it macroscopic friction coefficient} $S/P$.
In Figs. \ref{S_P_jam} and \ref{S_P_unjam},
we plot $S/P$ as a function of the density $\phi$
for both the {\it solid} and the {\it liquid branch}.
$S/P$ is almost independent of $\phi$ in the {\it solid branch},
while the apparent $\phi$-dependence of $S/P$ in the {\it liquid branch} 
is observed for the off-critical region $\phi - \phi_L(\mu) > 10^{-2}$,
which might be related to the discrepancy of $S$ and $P$ from the scaling law 
given by Eq. \eqref{liquid:eq} in Fig. \ref{unjam_scale}
because we assume that $S/P$ is independent of $\phi$ 
when we derive Eq. \eqref{liquid:eq} in Ref. \cite{Otsuki08}.
As shown in Fig. \ref{mu},
$S/P$ is almost independent 
of the friction coefficient $\mu$ except near $\mu=0$.
The $\mu$-dependence of  $S/P$ will be discussed elsewhere.

The existence of hysteresis is also known for frictionless 
granular particles under a
finite shear stress \cite{Ciamarra09-1}.
However, this hysteresis differs from that of our frictional system
shown in Fig. \ref{rheology}.
In Ref. \cite{Ciamarra09-1}, the stress is a control parameter,
and the states with $\dot \gamma =0$ and $\dot \gamma \neq 0$
are coexistent for a given shear stress.
On the other hand, the shear rate $\dot \gamma$ is a control parameter
in the SL method used for Fig. \ref{rheology}, 
and the two states for a given $\dot \gamma$
are those with large and small shear stress.

In conclusion,
we numerically investigate the sheared 
frictional granular particles, 
and find the existence of the hysteresis loops for different values of
the pressure and the shear stress, whose relevancy is numerically verified.
It is confirmed that
the critical densities which characterize the jamming transition
are split into three values,
where one of them is the true critical density for the jamming and the others are fictitious critical densities.
It is also verified that
the scaling relations \eqref{jam:eq} and \eqref{liquid:eq}
for frictionless particles can be used for the frictional systems
by using the fictitious critical densities.

\begin{acknowledgments}
We thank L. E. Silbert, T. Hatano, N. Mitarai, E. Brown and B. Tighe for their valuable discussions.
This work is partially supported by the 
Ministry of Education, Culture, Science and Technology (MEXT), Japan
 (Grant Nos. 21015016, 21540384, 21540388, and 22740260) and the Grant-in-Aid for the global COE program
"The Next Generation of Physics, Spun from Universality and Emergence"
from MEXT, Japan.
The numerical calculations were carried out on Altix3700 BX2 at 
the Yukawa Institute for Theoretical Physics (YITP), Kyoto University.
\end{acknowledgments}


\begin{thebibliography}{99}

\bibitem{Jaeger}
H. M. Jaeger, S. R. Nagel, and R. P. Behringer, Rev. Mod. Phys. {\bf 68},
1259 (1996).

\bibitem{Pusey} P. N. Pusey, in
\begin{em}Liquids, Freezing and the Glass Transition, Part II\end{em}, 
Les Houches Summer School Proceedings Vol. 51, edited
by J. -P. Hansen, D. Levesque, and J. Zinn-Justin (Elsevier, Amsterdam, 1991),
Chap. 10.

\bibitem{Durian} D. J. Durian and D. A. Weitz, 
\begin{em} "Foams," in Kirk-Othmer Encyclopedia of Chemical Technology, \end{em}
4th ed., edited by J. I. Kroschwitz (Wiley, New York,
1994), Vol. 11, p. 783.

\bibitem{Liu} A. J. Liu and S. R. Nagel, Nature {\bf 396}, 21 (1998). 

\bibitem{OHern02} C. S. O'Hern, S. A. Langer, A. J. Liu, and S. R. Nagel, Phys. Rev.
Lett. {\bf 88}, 075507 (2002). 

\bibitem{OHern03} C. S. O'Hern, L. E. Silbert, A. J. Liu, and S. R. Nagel, Phys. Rev. E {\bf 68}, 011306 (2003).



\bibitem{Olsson} P. Olsson and S. Teitel, Phys. Rev. Lett. \textbf{ 99}, 178001 (2007). 

\bibitem{Hatano07} T. Hatano, M. Otsuki, and S. Sasa, J. Phys. Soc. Jpn.
{\bf 76}, 023001 (2007). 

\bibitem{Hatano08} T. Hatano, 
J. Phys. Soc. Jpn. {\bf 77}, 123002 (2008).

\bibitem{Tighe} B. P. Tighe, E. Woldhuis, J. J. C. Remmers, W. van Saarloos,
and M. van Hecke,
Phys. Rev. Lett. {\bf 105}, 088303 (2010).

\bibitem{Hatano10} T. Hatano, 
Prog. Theor. Phys. Suppl. {\bf 184}, 143 (2010).


\bibitem{Otsuki08} M. Otsuki and H. Hayakawa,
Prog. Theor. Phys. {\bf 121}, 647 (2009).

\bibitem{Otsuki09} M. Otsuki and H. Hayakawa,
Phys. Rev. E {\bf 80}, 011308 (2009).

\bibitem{Otsuki10} M. Otsuki, H. Hayakawa, and S. Luding,
Prog. Theor. Phys. Suppl. {\bf 184}, 110 (2010).

\bibitem{Silbert02} L. E. Silbert, D. Ertas, G. S. Grest, T. C. Halsey, and
D. Levine, Phys. Rev. E {\bf 65}, 031304 (2002).

\bibitem{Zhang} H. P. Zhang and H. A. Makse, 
Phys. Rev. E {\bf 72}, 011301 (2005).

\bibitem{Shudyak} K. Shundyak, M. van Hecke, and W. van Saarloos, 
Phys. Rev. E {\bf 75}, 010301(R) (2007).

\bibitem{Somfai} E. Somfai, M. van Hecke, W. G. Ellenbroek, K. Shundyak,
and W. van Saarloos, Phys. Rev. E {\bf 75}, 020301(R) (2007).

\bibitem{Hecke} M. van Hecke, 
J. Phys.: Condens. Matter {\bf 22}, 033101 (2010).

\bibitem{Henkes}  
S. Henkes, M. van Hecke, and W. van Saarloos, Europhys. Lett.
{\bf 90}, 14003 (2010).

\bibitem{Silbert10} L. E. Silbert,
Soft Matter {\bf 6}, 2918 (2010).

\bibitem{Evans}
D. J. Evans and G. P. Morriss, {\it Statistical Mechanics of Nonequilibrium Liquids} 2nd ed.
(Cambridge University Press, Cambridge, 2008).


\bibitem{force:note}
We have checked that 
the value of the stress changed at most $10\%$
for $\mu = 0.0$ and $2.0$ if we adopt the normal force given as
$\bv{f}^{(n)'}_{ij} = h^{(n)}_{ij} \Theta(\sigma_i + \sigma_j - r_{ij}) 
\bv{n}_{ij}$.

\bibitem{DEM}
P. A. Cundall and O. D. L. Strack,
Geotechnique {\bf 29}, 47 (1979).


\bibitem{Hatano09}  
T. Hatano, Geophys. Res. Lett. {\bf 36}, L18304 (2009).

\bibitem{Heussinger09}
C. Heussinger and J.-L. Barrat,
Phys. Rev. Lett. {\bf 102}, 218303 (2009).

\bibitem{Vagberg10}
D. V{\aa}gberg, P. Olsson, and S. Teitel,
to be published in Phys. Rev. E (e-print arXiv:1007.2595).

\bibitem{Hysteresis} 
L. Vanel, D. Howell, D. Clark, R. P. Behringer, and E. Clement, 
Phys. Rev. E  {\bf 60}, R5040 (1999).

\bibitem{Eric}  
E. Brown and H. M. Jaeger,
Phys. Rev. Lett. {\bf 103}, 086001 (2009).

\bibitem{Ciamarra09}  
M. Pica Ciamarra, R. Pastore, M. Nicodemi, and A. Coniglio,
arXiv:0912.3140.

\bibitem{Somfai_v1} E. Somfai, M. van Hecke, W. G. Ellenbroek, K. Shundyak,
and W. van Saarloos, arXiv:0510506v1 (2005).

\bibitem{Olsson10} P. Olsson and S. Teitel, arXiv:1010.5885 (2010).

\bibitem{Wyart05} M. Wyart, L. E. Silbert, S. R. Nagel, and T. A Witten, 
Phys. Rev. E {\bf 72}, 051306 (2005). 

\bibitem{Ciamarra09-1}  
M. Pica Ciamarra and A. Coniglio,
Phys. Rev. Lett. {\bf 103}, 235701 (2009).

\end{thebibliography}
\end{document}